\documentclass[useAMS,usenatbib]{mn2e}

\usepackage{graphicx} 
\usepackage{epsfig}   
\usepackage{amsmath}
\usepackage{amssymb} 
\usepackage{mathrsfs}
\usepackage{url}
\usepackage{txfonts}
\usepackage{natbib}
\usepackage{units}
\usepackage{longtable, lscape, multirow}
\title[SPectro-Interferometric Data Analysis Software Tool]{\textit{SPIDAST}:~a new modular software to process spectro-interferometric measurements\thanks{Applied on observations made with ESO telescopes at the Paranal Observatory under Belgian VISA Guaranteed Time programme ID 083.D-029(A/B), 084.D-0131(A/B), 086.D-0067(A/B/C)} }
\author[P. Cruzal\`ebes et al.]{P. Cruzal\`ebes,$^{1}$\thanks{E-mail:
pierre.cruzalebes@oca.eu} Y. Rabbia,$^{1}$ A. Jorissen,$^{2}$ A. Spang,$^{1}$ S. Sacuto,$^{3,4}$ E. Pasquato,$^{2}$ \newauthor 
A. Chiavassa,$^{1,2}$  O. Chesneau$^{1}$ and P. Fr\'eville$^{5}$\\
$^{1}$Laboratoire Lagrange, UMR 7293, Universit\'e de Nice-Sophia Antipolis, CNRS, Observatoire de la C\^ote d'Azur, Bd de l'Observatoire, B.P. 4229, \\
F-06304 Nice cedex 4, France\\
$^{2}$Institut d'Astronomie et d'Astrophysique, Universit\'e Libre de Bruxelles, Campus Plaine C.P. 226, Bd du Triomphe, B-1050 Bruxelles, Belgium\\
$^{3}$Department of Astronomy and Space Physics, Uppsala Astronomical Observatory, Box 515, S-751 20 Uppsala, Sweden\\
$^{4}$Institute of Astronomy, University of Vienna, T\"urkenschanzstra\ss e 17, A-1180 Vienna, Austria\\
$^{5}$Club d'Astronomie Copernic, Maison des Associations, 642 rue des Batteries, F-83600 Fr\'ejus, France}
\begin{document}

\date{Version 2013 April 04}

\pagerange{\pageref{firstpage}--\pageref{lastpage}} \pubyear{tbd}

\maketitle

\label{firstpage}

\begin{abstract}
  {Extracting stellar fundamental parameters from SPectro-Interferometric (SPI) data requires reliable estimates of observables and with robust uncertainties (visibility, triple product, phase closure). A number of fine calibration procedures is necessary throughout the reduction process.
Testing departures from centro-symmetry of brightness distributions is a useful complement.
Developing  a set of automatic routines, called \textit{SPIDAST} (made available to the community) to reduce, calibrate and interpret raw data sets of instantaneous spectro-interferograms at the spectral channel level, we complement (and in some respects improve)  the ones contained in the \textit{amdlib} Data Reduction Software. Our new software \textit{SPIDAST} is designed to work in an automatic mode, free from subjective choices, while being versatile enough to suit various processing strategies. \textit{SPIDAST} performs  the following automated operations:~weighting of non-aberrant SPI data (visibility, triple product), fine spectral calibration (sub-pixel level), accurate and robust determinations of stellar diameters for calibrator sources (and their uncertainties as well), correction for the degradations of the interferometer response in visibility and triple product, calculation of the Centro-Symmetry Parameter (CSP) from the calibrated triple product, fit of 
parametric chromatic models on SPI observables, to extract model parameters.
\textit{SPIDAST} is currently applied to the scientific study of 18 cool giant and supergiant stars, observed with the VLTI/AMBER facility at medium resolution in the K band. Because part of their calibrators have no diameter in the current catalogs, \textit{SPIDAST} provides new determinations of the angular diameters of all calibrators. 
Comparison of \textit{SPIDAST} final calibrated observables with \textit{amdlib} determinations shows good agreement, under good and poor seeing conditions.}
\end{abstract}


\section{Introduction}

The power of optical-infrared interferometry to obtain information about the astronomical source morphology (including the angular size) is now well-established. To properly determine the source properties, the quality of the measurements is an issue, which is still the subject of active research. As with any other measuring apparatus, the absolute calibration of the instrument (including atmosphere) requires careful attention. Derived from the measurement of the mutual degree of coherence of the incident radiation field, on spatial frequencies sampled by the aperture-array configuration, the final interferometric observables are non-linear mixes of noisy quantities, and of parameter estimates with their own uncertainties. 

In this paper, we propose to revisit and extend the existing data processing and calibration methods, in the aim to obtain reliable estimates and robust uncertainties for calibrated measurements of visibility and complex triple product. The careful reduction process, described in the present paper, has been elaborated for the scientific study of a sample of 18 bright cool giant and supergiant stars (see Table~\ref{TabScience}). Measurements were obtained with the VLTI/AMBER facility at medium spectral resolution ($\mathscr{R}$=1500) in the $K$ band, using triplets of 1.80-m auxiliary telescopes. Observations have been conducted during 15 observing nights between May 2009 and December 2010 (under Belgian VISA Guaranteed Time). The aim of the present work is to test the data reduction and calibration software, on data with various qualities, that we started to develop in 2006.

   \begin{table}
  \caption{Science targets of our programme measured with VLTI/AMBER. Hipparcos parallaxes are in mas. Last column gives the calibrator(s) associated with each science target.} 
\label{TabScience}
\centering      
\begin{tabular}{lcccc}   
\hline                
 Name & Spec. Type  & $\varpi_\mathrm{Hip}$ & $m_{K}$ & Calibrator(s)\\
  \hline
   $\alpha$~Car & F0II & 10.6(6) & -1.3(3) & $\eta$~Col/$\iota$~Eri/HR~3282\\
   $\beta$~Cet & K0III & 33.9(2) & -0.3(4) & $\eta$~Cet\\
   $\alpha$~TrA & K2II & 8.4(2) & -1.2(1) & $\varepsilon$~TrA/$\gamma$~Lib/$o$~Sgr\\
   $\alpha$~Hya & K3II-III & 18.1(2) & -1.1(2) & $\lambda$~Hya\\
   $\zeta$~Ara & K3III & 6.7(2) & -0.6(2) & $\varepsilon$~TrA/$o$~Sgr\\
   $o_{1}$~CMa & K2.5Iab & 0.2(4) & 0.3(3) & HR~2411\\
   $\delta$~Oph & M0.5III & 19.1(2) & -1.2(2) & $\gamma$~Lib/$\varepsilon$~TrA\\
   $\gamma$~Hyi & M2III& 15.2(1) & -1.0(4) & $\alpha$~Ret \\     
   $o_{1}$~Ori & M3III & 5.0(23) & -0.7(2) & HR~2411\\
   $\sigma$~Lib & M3.5III & 11.3(3) & -1.4(2) & 51~Hya \\
   $\gamma$~Ret & M4III & 7.0(1) & -0.5(3) & $\alpha$~Ret\\
   L$_{2}$~Pup & M5IIIe & 15.6(10) & -1.8(1) & HR~3282/$\eta$~Col\\
   CE~Tau & M2Iab-b & 1.8(3) & -0.9(2) & 40~Ori\\
   T~Cet & M5.5Ib/II & 3.7(5) & -0.8(3) & $\gamma$~Scl/$\iota$~Eri\\
   TX~Psc & C7,2(N0)(Tc) & 3.6(4) & -0.5(3) & $\theta$~Psc\\
   R~Scl & C6,5ea(Np) & 2.1(15) & -0.1(1) & $\iota$~Eri\\
   W~Ori & C5,4(N5) & 2.6(10) & -0.5(4) & HR~2113/40~Ori\\
   TW~Oph & C5,5(Nb) & 3.7(12) & 0.5(4) & $o$~Sgr/$\gamma$~Lib\\     
  \hline
\end{tabular}
\end{table}

\section{Defining the basic interferometric observables} \label{SecObserv}

The \textit{coherent flux} $c_{ij}$ is provided by the measurement of the instantaneous observables delivered by the AMBER Data Reduction Software \textit{amdlib}\footnote{\url{www.mariotti.fr/data_processing_amber.htm}}. This quantity traces the sine-like molulated component superimposed on the continuum component, in the observed intensity distribution. It is computed using a $\chi^{2}$ linear fit of the individual interferograms in the detector plane, for each spectral channel \citep[see][for details]{tatulli07}. 

At this point, we define the other interferometric observables:
\begin{enumerate}
	\item the \textit{squared flux}, which is the product of the photometric fluxes $p_{i}$ and $p_{j}$, associated with the baseline $\vec{B}_{ij}$ 
     \begin{equation}
     f^{2}_{ij} = p_{i} p_{j}; 
     \label{InstantFlux2Def} 
   \end{equation}
  \item the \textit{squared visibility}, which is the ratio of the squared modulus of the coherent flux to the squared flux 
     \begin{equation}
     v^{2}_{ij} = \frac{\left| c_{ij} \right|^{2}}{f^{2}_{ij}};
     \label{InstantVis2Def} 
   \end{equation}
  \item the \textit{complex bispectrum} \citep{weigelt77}, which is the product of the 3 coherent fluxes contributing to the baseline triplet $\left(\vec{B}_{12}, \vec{B}_{23}, \vec{B}_{31}\right)$ 
     \begin{equation}
     b_{123} = c_{12}c_{23}c_{31}; 
     \label{InstantBispDef} 
   \end{equation}  
  \item the \textit{triple product}, which is the ratio of the bispectrum to the product of the 3 fluxes contributing to the baseline triplet 
     \begin{equation}
     t_{123} = \frac{b_{123}}{f_{12}f_{23}f_{31}}.
     \label{InstantTripDef} 
   \end{equation}
  \item the \textit{closure phase}, which is the argument of the complex bispectrum  (or triple product)
     \begin{equation}
	\psi_{123} = \arg b_{123} = \arg t_{123},
	   \end{equation}

\end{enumerate} 

%
   \begin{figure}
   \centering
   \includegraphics[scale=0.48, clip=true, trim=0 0 0 -25]{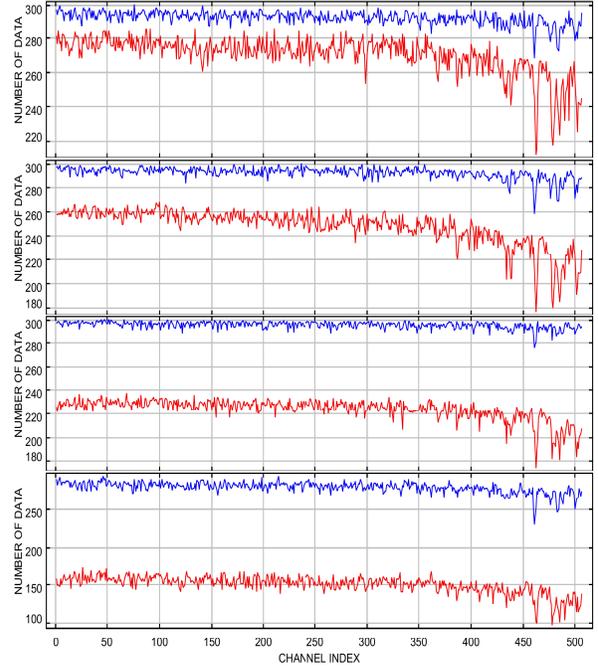}
    \caption{Spectral distribution  (in $K$) of the number of non-aberrant data, remaining after box-plot filtering, from an exposure of 300 input frames with the calibrator $\gamma$~Lib. From top to bottom:~baselines A0--D0 (\unit[32]{m}), D0--H0 (\unit[64]{m}), A0--H0 (\unit[96]{m}), and triplet A0--D0--H0. Blue lines:~good seeing conditions (seeing~angle=\unit[0.5]{\arcsec}, coherence~time=\unit[13]{ms}). Red lines:~poor seeing (\unit[1.7]{\arcsec}, \unit[1.3]{ms}).}
              \label{FigNbFr}%
    \end{figure}
%

\section{Reducing the raw data} \label{SecReduc}

To process the temporal series of frames (composing the ``exposures'') produced by the AMBER instrument, we use the data reduction software \textit{amdlib}, which consists of a core library of \textit{C} functions plus a high-level interface in the form of a \textit{Yorick}\footnote{\url{sourceforge.net/projects/yorick/}} plugin (a high-level language environment, similar to \textit{IDL} or \textit{Matlab}, which is in the public domain). 

The standard procedure to extract the basic interferometric observables from the AMBER raw data includes a frame-selection step, which provides a set of ``good'' frames according to a given criterion, or a sequence of several selection criteria \citep{malbet11}. The criterion used the most is based on the Signal to Noise Ratio (SNR) of the fringe-contrast in each frame. Two selection methods are provided in standard by \textit{amdlib}:
\begin{enumerate}
	\item retaining a given percentage of frames sorted according to their SNR values (called the ``percentage'' method),
	\item retaining the frames with SNR values higher than a given threshold (called the ``threshold'' method).
\end{enumerate}
With both methods, the choice of the optimal value (assumed not biasing the final results) of the selection criterion (in percentage of threshold) is done \textit{a posteriori}, from a sequence of processings using different values of the selection criterion \citep[see][for details]{millour07}. Two drawbacks of this are:~the need for a several-steps procedure (to be performed manually), and the risk for biasing the final results if the selection criterion is ill-determined. 

In our method, that we have implemented in a specific \textit{Yorick} script added to \textit{amdlib}, we apply the automatic procedure: 
\begin{enumerate}
	\item remove only the aberrant measurements (in squared visibility and triple product), \textit{for each spectral channel independently}. Using this specific script results in amounts of useful data larger than the amount obtained with the standard procedure, without increasing biasing effect (only the lower and upper tails of the input data histograms are rejected);
	\item assign a weight to each individual measurement. After many preliminary tests, we decided to choose weights based on the SNR of the coherent flux, and on the piston deviation;
	\item compute the temporal weighted average of the squared visibility and triple product  for each spectral channel, with uncertainties derived from the bootstrap technique invented by \citet{efron79}. 
\end{enumerate}

These automatic operations mentioned above, needing neither several-step selection nor manual choice of the optimal value of the selection criterion, are described in more details in the following subsections.
 
%
   \begin{figure}
   \centering
   \includegraphics[scale=0.5, clip=true, trim=0 0 0 -15]{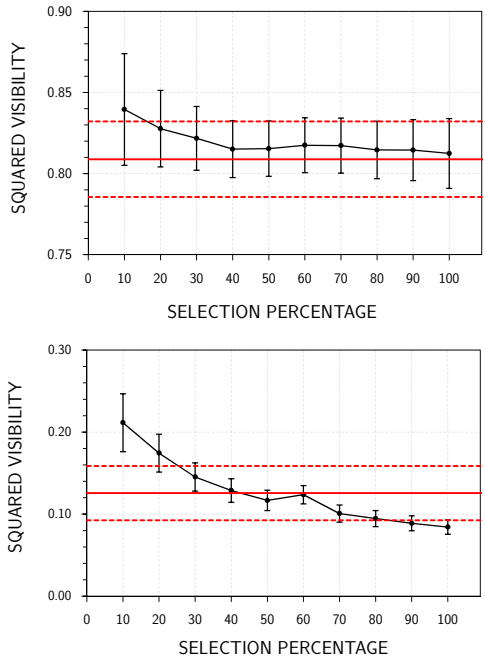}
   \caption{Squared visibility at \unit[2.2]{$\mu$m} obtained with \textit{amdlib} (black solid lines + error bars) w.r.t. the percentage of selected frames sorted according to their SNR value, for the calibrator $\gamma$~Lib (baseline A0--D0). Top panel:~good seeing conditions. Bottom panel:~bad seeing. Red lines:~values obtained with our method (solid line:~central value; dashed lines:~upper and lower limits given by the associated uncertainty).}%
              \label{FigComparoSelect}%
    \end{figure}
%

\subsection{Rejecting the aberrant data} \label{SubSecSelec}

To remove the aberrant measurements, we use a combination of physical and statistical basic criteria:
\begin{enumerate}
	\item strictly positive sum and product of the fluxes in the photometric channels, since negative flux measurements result from shortcomings in the determination of the continuum in each spectrum;
	\item coherent-flux SNR higher than unity, since inclusion of data with low SNR values reduces the reliability of the estimators of the interferometric observables \citep{millour08};
	\item squared visibility greater than $Q1-1.5\times IQR$ and lower than $Q3+1.5\times IQR$ (``box-plot filtering''), where $Q1$ denotes the first quartile, $Q3$ the third quartile, and $IQR=(Q3-Q1)$ the interquartile range \citep{hoaglin83}. With triple product data, this third criterion applies to the squared visibility for each baseline, and box-plot filtering on $\tan \psi_{123}$ is added to the list of criteria.
\end{enumerate}

%
   \begin{figure}
   \centering
   \includegraphics[scale=0.47, clip=true]{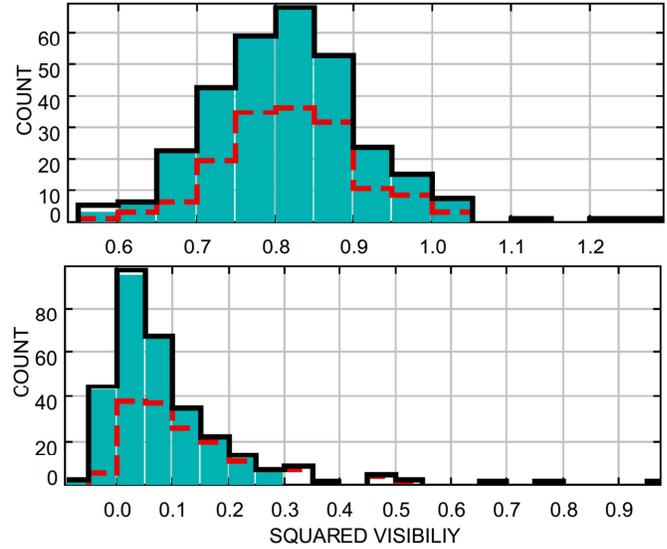}
    \caption{Histograms of the squared visibilities at \unit[2.2]{$\mu$m}, obtained with the calibrator $\gamma$~Lib (baseline A0--D0), under good (top panel) and poor (bottom panel) seeing conditions. Filled turquoise bars:~final histogram after aberrant-data rejection (our method). Full black steps:~initial histogram (no selection). Dashed red steps:~final histogram after 50\%-frame-selection with \textit{amdlib}.}
              \label{FigHisto}%
    \end{figure}
%

Based on the removal of aberrant measurements at the spectral channel level, rather than on the selection of the ``best'' frames, our approach provides an amount of useful data larger than the amount obtained with the standard frame-selection procedure. For data shown in Fig.~\ref{FigNbFr}, obtained for a ``good'' seeing   on the calibrator target $\gamma$~Lib, the number of removed input data varies only slightly from one spectral channel to the other (typically less than 3\%).
Besides, various amount of data are rejected, according to the seeing conditions.
 
Using the seeing parameters $\varepsilon_{0}$ (seeing angle) and $\tau_{0}$ (coherence time), the percentages of rejected data, respective to three baselines and two seeing conditions are given in the following lines (where ``good seeing'' refers to $\varepsilon_{0}$=\unit[0.5]{\arcsec}, $\tau_{0}$=\unit[13]{ms}, and ``poor seeing'' refers to $\varepsilon_{0}$=\unit[1.7]{\arcsec}, $\tau_{0}$=\unit[1.3] {ms}):
\begin{itemize}
	\item shortest baseline A0--D0 (\unit[32]{m}):~2\% (good seeing), to 9\% (poor seeing);
	\item intermediate baseline D0--H0 (\unit[64]{m}):~2\% to 16\%;
  \item longest baseline A0--H0 (\unit[96]{m}):~2\% to 25\%;
	\item in addition, for the baseline triplet A0--D0--H0, the rejection of data amounts 7\% and 50\%.
\end{itemize}

%
   \begin{figure*}
   \centering
   \includegraphics[scale=0.5, clip=true, trim=0 0 0 -5]{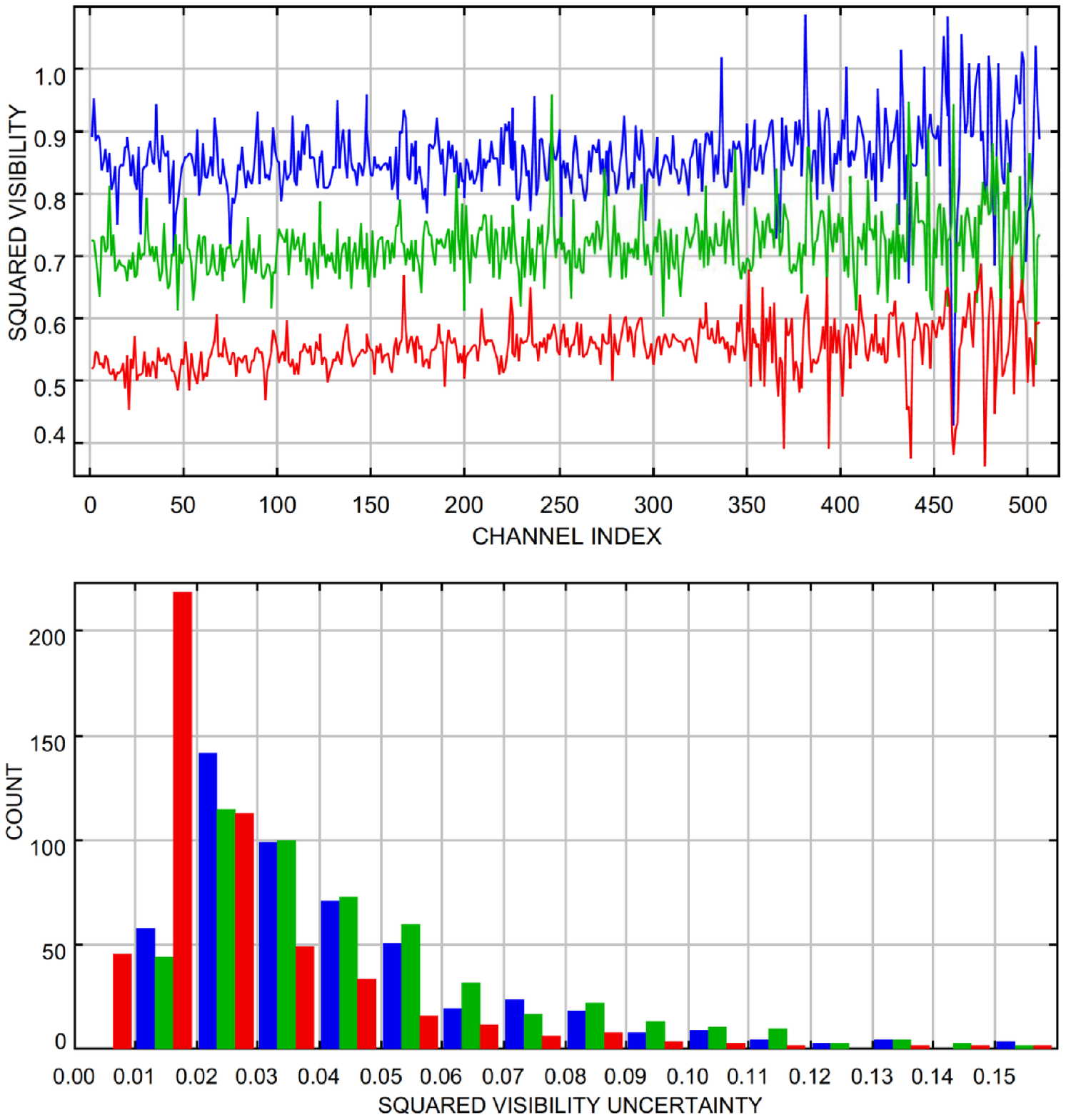}
   \includegraphics[scale=0.5, clip=true, trim=0 0 0 -5]{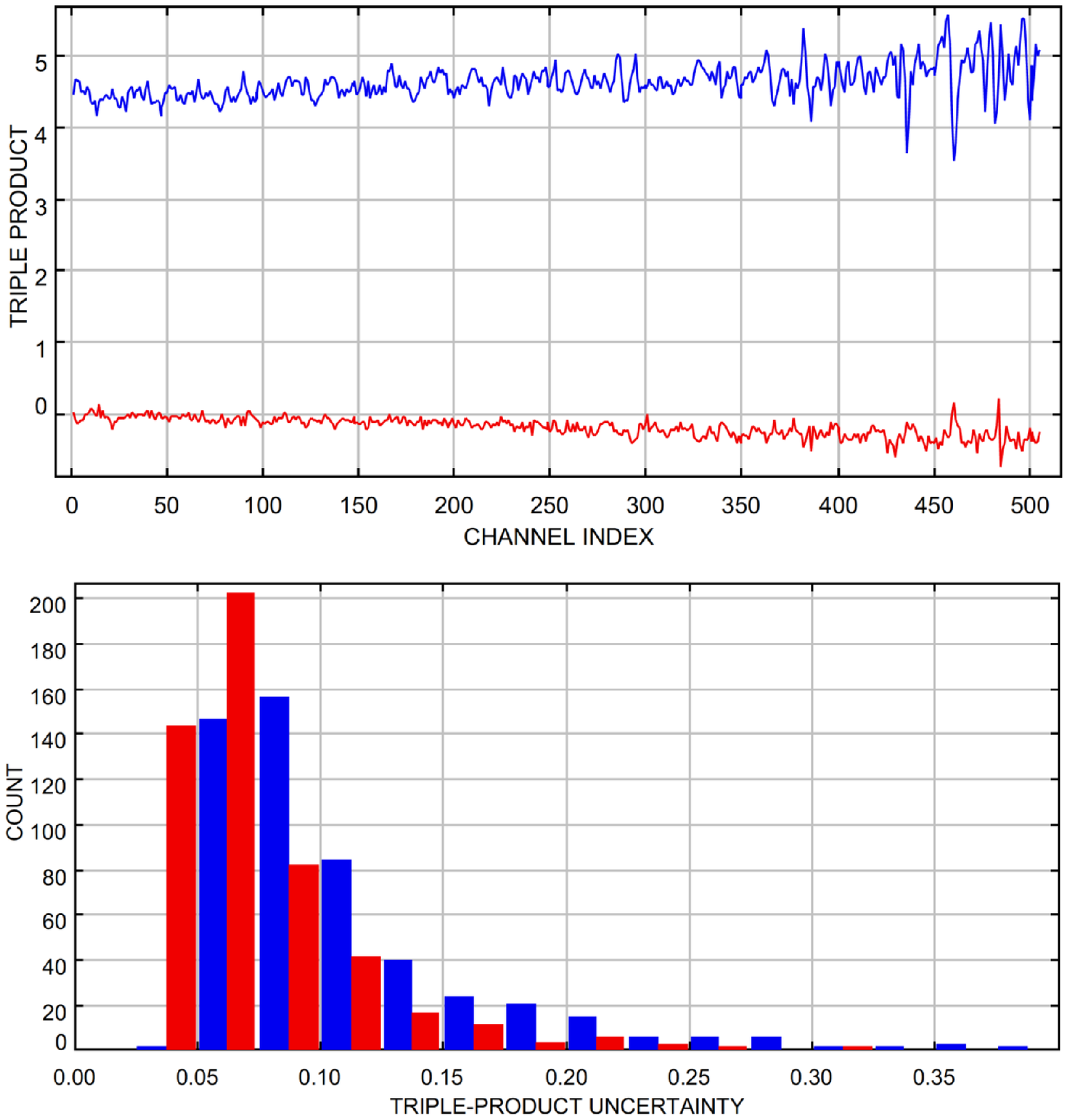}
   \caption{Interferometric observables given by our reduction code with the calibrator $\gamma$~Lib, before calibration (good seeing). Top left panel:~squared visibilities for the baselines A0--D0 (blue), D0--H0 (green), and A0--H0 (red) w.r.t. the channel index in $K$. Top right panel:~real (blue) and imaginary (red) parts of the triple product for the baseline triplet A0--D0--H0. Because of their low levels, uncertainties in the bottom panels are drawn using histograms, rather than error bars (with the same colour code as the top panels).}
              \label{FigGamLibRaw}%
    \end{figure*}
%

Figure~\ref{FigComparoSelect} shows the variation of the final squared visibility (at \unit[2.2]{$\mu$m}) produced with  \textit{amdlib}, w.r.t. the percentage of selected frames, based on the SNR and pertaining to the calibrator target $\gamma$~Lib, observed with the shortest baseline A0--D0 (\unit[32]{m}) in good and poor seeing conditions (top panel and bottom panel respectively).
The squared visibility produced with our method is shown for comparison. 

In this example, the position of the plateau of each curve indicates a value of the selection threshold of 50\%, giving the \textit{a posteriori} determination with \textit{amdlib}. There we find that our automatic method gives a squared-visibility value close to the 50\% value, without a need for the several-steps procedure, in good and poor seeing conditions. Besides, Fig.~\ref{FigHisto} shows that the standard \textit{amdlib} frame-selection method keeps the extremal V$^{2}$ values in poor seeing condition, which may bias the result of the calculation of the average over the frames contained in each exposure. On the contrary, the box-plot filtering used with our method rejects only the lower and upper tails of the V$^{2}$ histogram, thus leading to estimates of the temporal average, more reliable than with the frame-selection method. 

Next subsection describes the weighting scheme used to average the non-aberrant data over each exposure. Note that selecting the frames with the standard \textit{amdlib} procedure is equivalent to assigning unity weights to the selected frames, and null weights to the others. 

\subsection{Computing the raw observables} \label{SubSecAver}

To reduce the instrumental effects on the data, \textit{at sthe pectral channel level}, we compute in each exposure (300 frames in our example) the weighted average of the squared visibility, and of the complex triple product. For the visibility, as weight associated with each frame we use the ratio of the SNR to the relative excursion of the piston, where the relative excursion of the piston itself is the ratio of the piston excursion (absolute value) for a given frame, to the average over the whole exposure. For the triple product, each individual weight is given by the geometrical mean of the three weights associated with the three baselines. The final uncertainties are derived from the variance of the distribution of the weighted means, obtained by random sampling with replacement of the original series of data (direct-bootstrap method) \citep{efron93}. Note that our specific script computes the real and imaginary parts of the triple product, from which we extract the calibrated closure phase (see Sect.~\ref{SubSecIntercal}).

Figure~\ref{FigGamLibRaw} shows an example of squared visibility and triple product produced by the reduction script (before calibration) from one single exposure of 300 input frames, obtained with the calibrator target $\gamma$~Lib, for the three baselines A0--D0 (\unit[32]{m}), D0--H0 (\unit[64]{m}), and A0--H0 (\unit[96]{m}). 
 
In complement to these interferometric observables, we compute two other quantities used further in the calibration procedure (see Sect.~\ref{SubSecRespcorr}):
\begin{enumerate}
  \item the weight of each exposure, defined as the product of two ratios:~one is the average of the SNR to its variance; the other is the average of the inverse of the piston excursion (absolute value) to its variance;
	\item the final number of non-aberrant data, for each spectral channel.
\end{enumerate}

\section{Calibrating the data} \label{SecCalib}

The calibration process is the key point to obtain accurate estimates of the ``true'' (intrinsic) observables of the science targets. To correct the measurements for environmental and instrumental instabilities, we observe reference stars (the calibrators), with known angular diameters and independently well-defined brightness distributions. 

To calibrate the wavelength-dependent measurements obtained with VLTI/AMBER, we have developed a library of IDL functions, included in the \textit{SPIDAST} (SPectro-Interferometric Data Analysis Software Tool) modular software suite \citep{cruzalebes08,cruzalebes10}, which allows one to:
\begin{enumerate}
\item link each spectral channel to a wavelength value. We use a specific method based on the correlation of calibrator measured spectra with synthetic templates given by the MARCS model \citep{gustafsson08};
\item measure and correct for the degradation of the spatial coherence. As calibrators, we use stars with brightness distribution of Limb-Darkened (LD) discs, for which angular diameters are given by fits of synthetic spectra on the wide-band spectrophotometric measurements in the infrared \citep{cruzalebes10}.
\end{enumerate}

%
   \begin{figure}
   \centering
   \includegraphics[scale=0.5, clip=true, trim=0 0 0 -5]{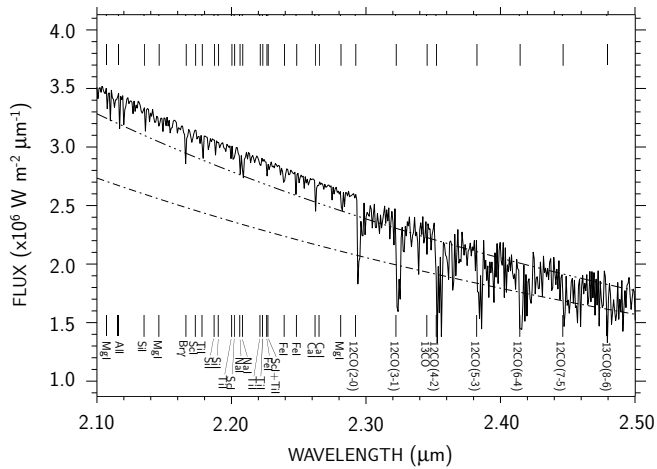}
   \caption{Spectral distribution of the synthetic model surface flux of $\gamma$~Lib, produced by the MARCS + Turbospectrum code in $K$ . Dash-dot line: black-body continuous spectrum with $T_\mathrm{eff}$=\unit[4660]{K}. Dash-dot-dot line : Engelke continuous spectrum with the same effective temperature.}
              \label{FigGamLibSpec}%
    \end{figure}
%

\subsection{Computing the spectral shifts} \label{SubSecWavecal}

To reach a high precision in the angular diameter estimation using model fitting, spectral calibration is a critical point \citep[see e.g.,][]{wittkowski08,domiciano08,stefl11}. Given the lack of any internal instrumental module for wavelength calibration in the optical setup of AMBER \citep{robbe07,petrov07}, \textit{amdlib} provides calibrated wavelength tables, computed from a theoretical polynomial dispersion law \citep{merand10}, but with coefficients still badly known, leading to wavelength shifts up to \unit[$\sim$10]{pixels} in the detector plane, at medium spectral resolution \citep{malbet11}. To correct for this drawback, \textit{amdlib} shifts the wavelength table using the correlation of the measured spectrum with a template table containing the telluric lines.

To improve the precision of the wavelength calibration, observers usually compute the coefficients of the dispersion law by identifying visually some prominent telluric lines in the measured spectrum \citep[e.g.,][]{kraus09,weigelt11,ohnaka11}. We propose an alternative automatic approach, based on the cross-correlation of the calibrator measured spectra with synthetic templates produced by the MARCS + Turbospectrum codes \citep{alvarez98,plez12}, including the telluric lines as well. This method was previously suggested by \citet{plez03b} for Gaia, and by \citet{decin04b} or \citet{decin07} to calibrate the SPITZER spectrograms. 
It is particularly suitable for stellar spectra showing easily identifiable spectral features, as it is the case with our sample of cool giant calibrators, showing strong CO~bands in the \unit[2.126--2.474]{$\mu$m} spectral range \citep{marti11}. Figure~\ref{FigGamLibSpec} shows the synthetic spectrum produced by the MARCS + Turbospectrum code, for the calibrator $\gamma$~Lib. Some reference spectral lines are shown for identification of the corresponding stellar atmospheric elements.

%
   \begin{figure}
   \centering
   \includegraphics[scale=0.47, clip=true, trim=0 0 0 -5]{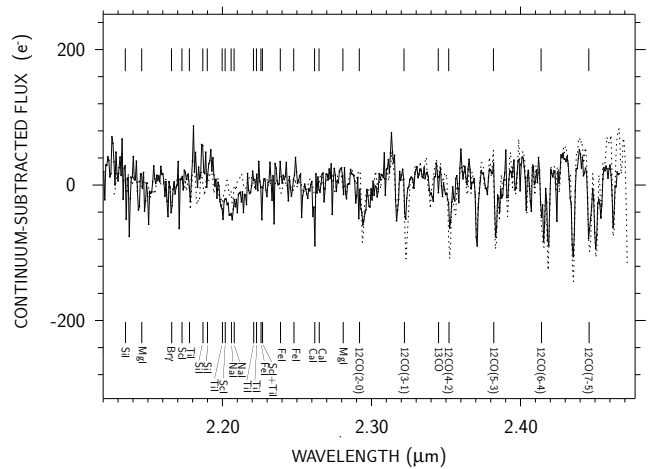}
   \caption{Spectral distribution of the continuum-subtracted flux of $\gamma$~Lib measured with VLTI/AMBER, after spectral calibration. Dashed line:~model spectrum (including tellurics).}%
              \label{FigCalibSpec}%
    \end{figure}
%

%
   \begin{figure}
   \centering
   \includegraphics[scale=0.43, clip=true, trim=0 0 0 -5]{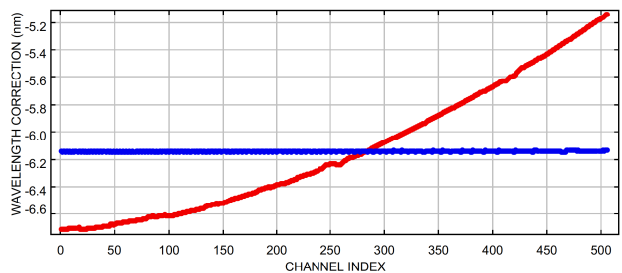}
   \caption{Wavelength correction from the initial polynomial dispersion law of AMBER, obtained with \textit{SPIDAST} (in red) and with \textit{amdlib} (in blue), for each spectral channel of the $K$ band (first observing night).}
              \label{FigComparoWave}%
    \end{figure}
%

For each exposure, obtained with each calibrator, and with the working instrumental setting, our automatic spectral calibration process contains the following steps:
\begin{enumerate}
	\item apply the heliocentric and systemic radial velocity corrections, and multiply the synthetic spectrum by the atmospheric and instrumental transmittance profiles. Atmospheric transmission data for the southern sites (CTIO, Chile) are given by the USAF atmospheric code PLEXUS \citep*{cohen03}; 
	\item remove the continuum parts of the raw and synthetic spectra, and normalize the resulting spectra. As a first approximation, we estimate the continuum by decreasing the spectral resolution to $\mathscr{R}$=40. The main drawback of this method is to produce apparent ``pseudo-continua'' that are lower than the real continuum levels, even in almost line-free regions \citep{rix04}. Since our goal is only to perform spectral calibration by correlation, this drawback has no effect on the final result; 
	\item divide the spectrum in contiguous sub-windows of the same size, and compute the wavelength shift by phase-correlation \citep{vera08} in each sub-window. If the spectrum is divided in $n$ sub-windows, we apply a polynomial law of degree $n-1$ to correct for the wavelength shifts. Sub-pixel precision is obtained by embedding the cross-power spectrum in the middle of a two-times larger array of zeros, before computing the phase-correlation function, by inverse Fourier Transform of the final cross-power spectrum \citep*{guizar08};
	\item finds the position of the barycenter of the correlation peak for each sub-window, computes the coefficients of the interpolation law (assumed to be polynomial), and corrects the global wavelength table;
	\item calculates the residuals of the wavelength shifts, using once again the phase-correlation method on the wavelength-corrected spectrum.
\end{enumerate}

   \begin{table}
  \caption{Angular diameter values (in mas) of the suitable calibrators, found in the CCSL, derived from the final fit of the MARCS-model spectrum, and found in the JSDC. The third column gives the effective temperature (in K) deduced from the spectral type, used for the model.} 
\label{TabDiamCal}
\centering      
\begin{tabular}{lccccc}   
\hline                
 Name & Spec. Type & $T_\mathrm{eff}$ & $\phi_\mathrm{CCSL}$ & $\phi_\mathrm{final}$ & $\phi_\mathrm{JSDC}$\\
  \hline
   $\alpha$~Ret & G8II-III & 4780(230) & - & 2.54$\left(^{4}_{9}\right)$ & 2.5(2)\\
  $\varphi_{2}$~Ori & K0IIIb & 4670(230) & 2.20(2) & 2.263$\left(^{6}_{5}\right)$ & 2.1(2)\\
   $\eta$~Col & K0III & 4660(230) & 2.48(3) & 2.38$\left(^{7}_{8}\right)$ & 2.7(2)\\
   $\lambda$~Hya & K0III & 4660(230) & - & 2.62$\left(^{14}_{29}\right)$ & -\\
   $\gamma$~Lib & K0III & 4660(230) & - & 2.31$\left(^{7}_{15}\right)$ & -\\
   $o$~Sgr & K0III & 4660(230) & - & 2.50$\left(^{12}_{27}\right)$ & 2.4(2)\\
   $\iota$~Eri & K0.5IIIb & 4600(220) & 2.18(2) & 2.35$\left(^{4}_{37}\right)$ & 2.5(2)\\
   $\theta$~Psc & K0.5III & 4580(220) & 2.00(2) & 2.11$\left(^{2}_{24}\right)$ & 2.1(2)\\
   $\gamma$~Scl & K1III & 4510(220) & 2.13(3) & 2.039$\left(^{14}_{7}\right)$ & 2.2(2)\\
   HR~2113 & K1.5III & 4440(220) & - & 2.48(2) & 2.4(2)\\
   $\varepsilon$~TrA & K1.5III & 4440(220) & 2.56(7) & 2.43(2) & -\\
   $\eta$~Cet & K1.5III & 4440(220) & 3.44(4) & 3.323$\left(^{28}_{9}\right)$ & 3.3(2)\\
   HR~3282 & K2.5II-III & 4330(210) & 2.54(4) & 2.32(7) & -\\
   HR~2411 & K3III & 4260(210) & 1.90(3) & 1.76$\left(^{1}_{15}\right)$ & -\\     
   51~Hya & K3III & 4260(210) & 2.28(3) & 2.23$\left(^{7}_{19}\right)$ & 2.4(2)\\
  \hline
\end{tabular}
\end{table}

For each observing night, the final wavelength table, associated with the working instrumental setting, is obtained by ensemble average of the corrected tables (of all exposures with all calibrators measured with the same instrumental setting), rejecting the tables showing wavelength-shift residuals higher than \unit[8]{nm} ($\sim$5 times the nominal spectral resolution).
Figure~\ref{FigCalibSpec} shows the final continuum-corrected spectrum, given by the spectral-calibration procedure, with the calibrator $\gamma$~Lib. The model spectrum drawn with the dashed line is produced by the MARCS + Turbospectrum code. 

To show the difference between the wavelength-calibration provided by \textit{SPIDAST} and by \textit{amdlib}, we plot in Fig.~\ref{FigComparoWave} the corrections from the initial polynomial dispersion law of AMBER, given by the two softwares. While \textit{amdlib} computes only a unique correction value applied over the whole spectrum, our method computes a polynomial correction (here parabolic from three sub-windows), with sub-pixel deviations with respect to the polynomial law, caused by the final ensemble averaging of the corrected wavelength tables. The choice of the degree of the polynomial correction provided by \textit{SPIDAST} allows the accuracy of the wavelength correction to be adapted to the scientific programme, reaching a sub-pixel precision if necessary.

%
   \begin{figure}
   \centering
   \includegraphics[scale=0.5, clip=true, trim=0 0 0 -5]{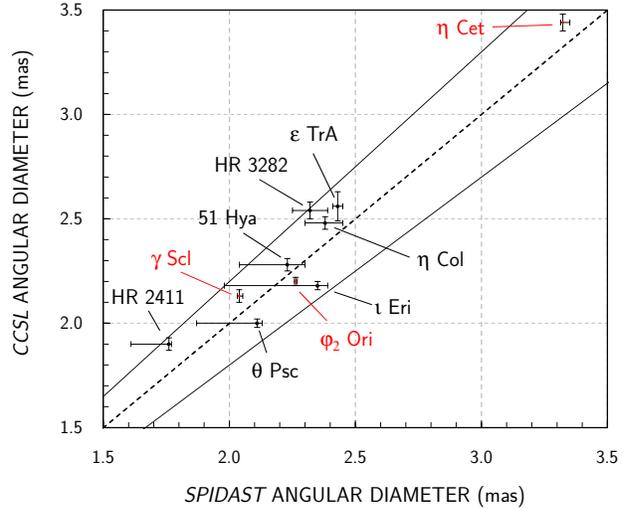}
   \caption{Value of the angular diameter of each calibrator of our observing sample found in the CCSL, versus the final value deduced from \textit{SPIDAST}, fitting the MARCS-model spectra on calibrated templates of \citet{cohen99} (red dots), and on IRAS-LRS measurements (black dots). Solid lines:~$\pm$~10\% thresholds.}%
              \label{FigDiamCata}%
    \end{figure}
%

\subsection{Calibrating spectro-interferometric data} \label{SubSecIntercal}

In order to assess and correct for the measurement defects on the science targets, proper interferometric calibration (not yet fully supported with \textit{amdlib}) is needed. The \textit{Instrument Response Function} (IRF), also called ``transfer function'' \citep{perrin03,vbelle05,boden07,cruzalebes10}, is derived from observations of calibrators as concomitant as possible to the ones of the science targets. 

The standard calibration method is based on the assumption that a given ``raw'' (measured) quantity $q$ of any observable $\mathscr{Q}$, is equal to the product of the ``true'' (intrinsic) quantity $\mathcal{Q}$, multiplied by a global degradation factor $R_\mathscr{Q}$ (in other words the IRF). Such a factor cannot practically be modellised with sufficient reliability (numerous parameters and some of them remaining unknown). 

To calibrate the quantity related to the science target, a second assumption applies, as soon as appropriate conditions (described later on)  are satisfied and can be then determined by using the relation
\begin{equation}
     \frac{q_\mathrm{sci}}{\mathcal{Q}_\mathrm{sci}} = \frac{q_\mathrm{cal}}{\mathcal{Q}_\mathrm{cal}} = R_\mathscr{Q},
     \label{TrueObs} 
   \end{equation}	
with self explanatory notations, stating that the degradation factor is identical for both the calibration source and the science target. This assertion is all the more valid that the required conditions are fullfilled. The relation here above yields an empirical determination  of $\mathcal{Q}_\mathrm{sci}$, via $R_\mathscr{Q}$, relying on commonplace statistical tools. The main condition to satisfy is that the  IRF is stable enough, and the standard approach to avoid too large variations of the IRF is to perform sequences, where observation of calibrator and science targets are interlaced and repeated several times, rapidly sampling the IRF. So, the actual stability requirement is that the IRF might be slowly variable over a given ``calibrator-science-calibrator'' observation sequence.   
Besides, the calibrator and the science target should be close enough regarding  time of observation, angular separation, and brightness (in the spectral domain of work), so that adjustments of the whole interferometer would not be significantly modified  from one source to the other. In these conditions, only the atmospheric turbulence might cause substantial degradations of the measurements. For this latter case, servo-loop systems are to be used to reduce turbulence effects. 

Usually, calibrators are found in dedicated catalogs and the corresponding sources should meet the desired requirements. However, the number of calibrators is necessarily limited and somewhat depends on the type of science targets, so that the selected calibrators might stand out from the ideal conditions:~point-like source, small angular separation, brightness matching. 

In practice, the limited choice of calibrators makes it necessary to accept ``partially'' resolved targets \citep{vbelle05}, angular separations counted in degrees, and brightness mismatch amounting some units of magnitudes. For example, a bright calibrator is rarely found in a close angular vicinity of a given bright science target. Such a situation nevertheless remains convenient for the calibration procedure, since switching between science targets and calibrators is fast enough and does not affect the configuration of the interferometer, and its response as well.  

This procedure starts with a first selection of calibrators via the \textit{SearchCal} tool\footnote{\url{www.mariotti.fr/searchcal_page.htm}}, created by the JMMC working group ``Catalogue of calibration sources'', providing rough estimates  for angular diameters, with comparatively large uncertainties. Following this first selection, the sources are searched in the calibrator catalogues of \citet{borde02} and \citet{merand05}, in order to find angular diameters of better quality. Since some calibrators have not a diameter in the published catalogues, it is necessary to make our own determinations of those diameters (see Subsect.~\ref{SubSecCaldiam}). Moreover,  in order to control these determinations and to build a homogeneous set of  diameters, we apply our own method to recalculate all the diameters for the selected calibrators. The good agreement found between our determinations and the ones of the Bord\'e's Catalogue of Calibrator Stars for Long-Baseline Stellar Interferometry (CCSL) (as shown in Table~\ref{TabDiamCal}, and in Fig.~\ref{FigDiamCata}), attests to the reliability of our specific determinations of angular diameters.

%
   \begin{figure}
   \centering
   \includegraphics[scale=0.5, clip=true, trim=0 0 0 -20]{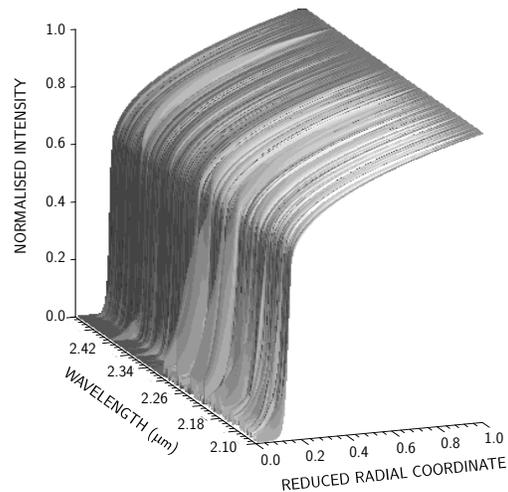}
   \caption{$\lambda$--$\mu$ map of CLV profiles produced by the MARCS + Turbospectrum code, in the $K$ band, for a K0III-type star.}%
              \label{FigIntens3d}%
    \end{figure}
%

\subsubsection{Determining calibrator angular diameters} \label{SubSecCaldiam}

To derive the IRF for each spectro-interferometric (SPI) observable, we firstly need to get a reliable estimate of the angular diameter of each calibrator, and its associated error. Fitting the stellar models of \citet{kurucz79} on the absolutely calibrated spectro-photometric templates of \citet{cohen99}, the CCSL contains 374 calibrators, with limb-darkened angular diameters ranging from \unit[1 to 3]{mas}. Since our selected calibrators might have no angular-diameter estimate (one third of the calibrators of our dataset have no value found in the catalogues), or poorly estimated, we have added to \textit{SPIDAST} various routines for determination of the angular diameter from indirect methods, presented in detail in \citet{cruzalebes10}. These routines compute the angular diameter: 
\begin{enumerate}
	\item by combining the linear diameter, deduced from absolute luminosity and effective temperature from the Morgan-Keenan (MK) spectral type, with the parallax;
	\item by using experimental laws based on the interstellar-corrected colour index (surface brightness method);
	\item by scaling synthetic spectra on broadband photometric measurements (infrared flux method);
	\item by fitting synthetic spectra on infrared spectro-photometric measurements (see Sect.~\ref{SecFit}). 
\end{enumerate}

%
   \begin{figure}
   \centering
   \includegraphics[scale=0.75, clip=true, trim=0 0 0 -5]{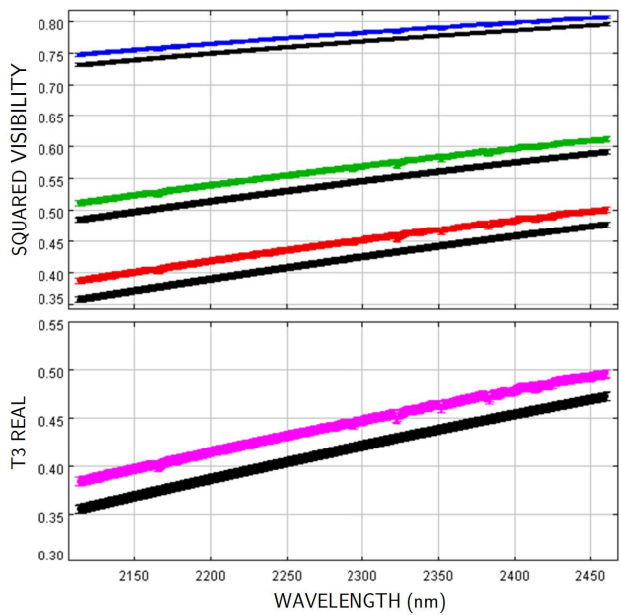}
   \caption{Theoretical interferometric observables of the calibrator $\eta$~Ceti, produced by the LD-MARCS model (in colour), and by the UD model with the same value of the angular diameter (in black). Top panel:~squared visibilities for the baselines G1--I1 (blue), D0--G1 (green), and D0--I1 (red). Bottom panel:~real part of the triple product produced by the models, for the baseline triplet D0--G1--I1. For each curve, the thickness gives the uncertainty.}
              \label{FigEtaCet}%
    \end{figure}
%

Table~\ref{TabDiamCal} gives the angular diameter values of our calibrators found in the CCSL, as well as our final estimates, derived from the fit of MARCS + Turbospectrum synthetic spectra on photometric or spectro-photometric measurements. In the last column, we also give the estimates found in the JMMC Stellar Diameter Catalogue (JSDC) of \citet{lafrasse10}, with accuracies between 7 and 10\% \citep{delfosse04,bonneau06}.

Figure~\ref{FigDiamCata} shows the comparison of our estimates with the CCSL values. The mean difference between our estimates and the CCSL is 5\%. The mean difference for the three targets with the available calibrated templates of Cohen (i.e. $\phi_{2}$~Ori, $\gamma$~Scl, and $\eta$~Cet) is lower: 3.5\%. This shows the satisfactory agreement between our results and the CCSL, which confirms the reliability of our approach for determination of the calibrator angular diameters, implemented in \textit{SPIDAST}.

%
   \begin{figure}
   \centering
   \includegraphics[scale=0.45, clip=true, trim=0 0 0 -15]{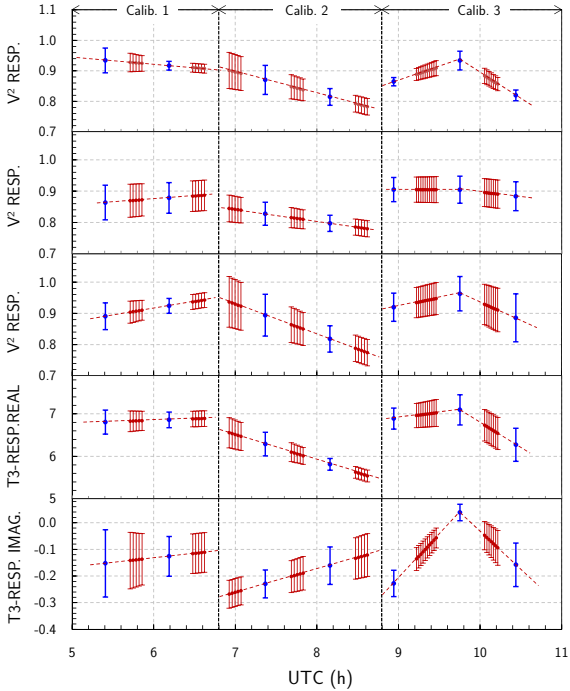}
   \caption{Temporal variation of the IRF ($\lambda$=\unit[2.2]{$\mu$m}) during one observing night. Blue dots: response values averaged over each OB, successively obtained with the calibrators $\gamma$~Lib, $\varepsilon$~TrA, and $o$~Sgr. Red dots:~response values interpolated at the time of observation of the science targets (red dashed lines:~interpolated response). From top to bottom:~response in squared visibility for the baselines A0--D0, D0--H0, A0--H0, real and imaginary parts of the response in triple product for the triplet A0--D0--H0.}
              \label{FigResp}%
    \end{figure}
%

\subsubsection{Modelling the calibrator visibility and triple product} \label{SubSecCalmod}

Once we determined the angular diameter of each calibrator, and its associated uncertainty, we derive the IRF for each SPI observable, computing the ratio of the observable measured on the calibrator, to the synthetic observable, assumed to represent the ``true'' calibrator observable. Each synthetic SPI observable is derived from the synthetic coherent flux $C_{\lambda}^\mathrm{model}$ 
\begin{enumerate}
	\item the synthetic flux (Eq.~\ref{InstantFlux2Def}) is given by the synthetic coherent flux computed for the null baseline (visibility unity);
	\item the squared synthetic visibility (Eq.~\ref{InstantVis2Def}) by the ratio of the squared synthetic coherent flux to the synthetic squared flux;
	\item the synthetic bispectrum (Eq.~\ref{InstantBispDef}) by the product of the synthetic coherent flux for the three baselines composing each triplet;
	\item and the synthetic triple product (Eq.~\ref{InstantTripDef}) by the ratio of the synthetic bispectrum to the cubed synthetic flux.
\end{enumerate}
Applying the van-Cittert-Zernike theorem, the model coherent flux of each calibrator, assumed to emit a luminous intensity with circular symmetry, is derived from the Hankel transform of the source intrinsic \textit{spectral radiance} $L_{\lambda}$ (in $\mathrm{W\:m^{-2}\:\mu m^{-1}\:sr^{-1}}$)
    \begin{equation}
   C_{\lambda}^\mathrm{model} = \pi \frac{\phi^{2}}{4} \int^{\infty}_{0} L_{\lambda}\left( r \right) \: 
   \mathrm{J}_{0}\left( \pi r \phi \frac{B_{ij}}{\lambda} \right) \: r dr ,
   \label{ModelCohLD} 
   \end{equation}
where $\phi$ is the stellar angular diameter, $r$ the impact parameter, i.e. the distance from the centre of the stellar disc ($r=0$ at disc center), $B_{ij}$ the length of the projected baseline of the (i,j)-pair of apertures, and J$_{0}$ the Bessel function of order zero.

Since compact photospheres are known to deviate from simple uniform discs \citep[see][and references therein]{hajian98}, radially decreasing from their photometric centre, we use the MARCS + Turbospectrum codes to produce models of Centre-to-Limb Variation (CLV) profiles, with input parameters derived from the MK spectral type. Figure~\ref{FigIntens3d} shows the $\lambda$--$\mu$ map of typical CLV profiles given by the model, where the reduced radial coordinate $\mu$ is given by $\mu$=$\sqrt{1-r^2}$ \citep{young03}. 

With ``partially'' resolved calibrators, the deviation from the UD model increases with the baseline length. Figure~\ref{FigEtaCet} shows the squared visibility and the real part of the triple product produced by the MARCS and the UD models, for the calibrator $\eta$~Ceti ($\phi_\mathrm{final}$=\unit[3.32$\pm$0.02]{mas}), deduced from measurements obtained with the D0--I1--G1 baseline triplet (G1--I1=\unit[46]{m}, D0--G1=\unit[69]{m}, and D0--I1=\unit[79]{m}). In this example, the differences in visibility and triple product between the UD model and the LD-MARCS model are larger than the uncertainty of the synthetic observables (derived by propagating the uncertainty on the angular diameter), hence the justification to use the LD model instead of the UD model.

   \begin{figure*}
	\centering
   \includegraphics[scale=0.95, clip=true, trim=0 0 0 -15]{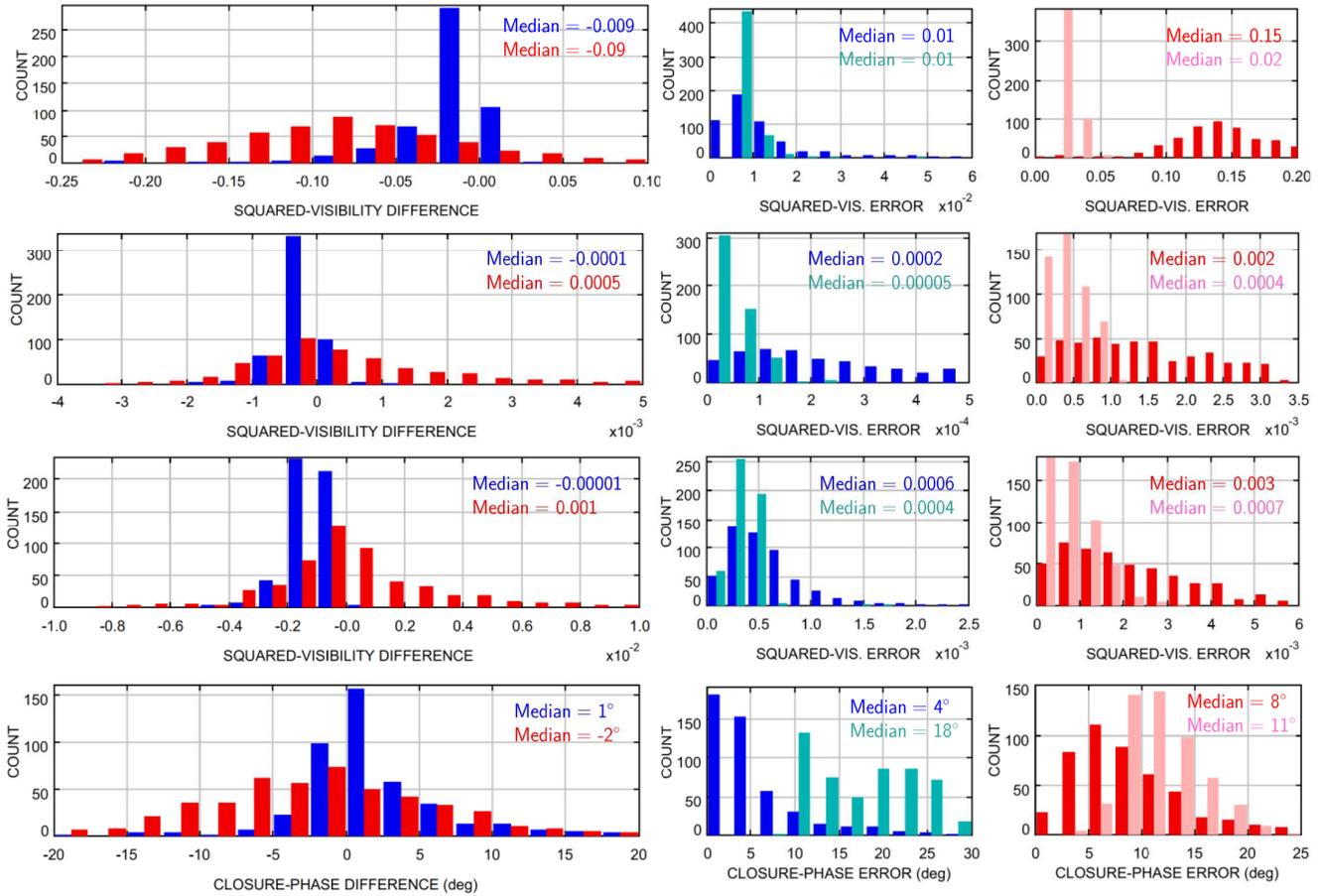}
  \caption{Histograms of calibrated quantities pertaining to the science target $\delta$~Oph, obtained in good (blue) and poor (red) seeing conditions. From top to bottom:~squared visibilities with the baselines A0--D0, D0--H0, A0--H0, and closure phase with the triplet A0--D0--H0. Long panels of the first column:~difference between \textit{SPIDAST} and \textit{amdlib}. Small panels of the other columns:~associated uncertainties with good (blue:~\textit{SPIDAST}, turquoise:~\textit{amdlib}), and poor seeing (red:~\textit{SPIDAST}, pink:~\textit{amdlib}). }
              \label{FigDelOphCalib}%
    \end{figure*}
%

\subsubsection{Correcting for the degradations of the IRF} \label{SubSecRespcorr}

The main source of non-stationary instabilities which affects the fringe formation is the atmospheric-phase turbulence \citep{roddier81}. This issue is partially solved thanks to the use of fringe-tracking servo-loops scanning the fringes with cycle periods smaller than the seeing coherence time, defined as the time over which the phase fluctuations remain coherent \citep{davis96,kellerer07}. Thus, the residual instabilities are expected to cause only slow drifts of the IRF between observations of the science targets and their associated calibrators. In order to estimate the response at the time when each science target is measured, linear interpolation between the successive measurements on its calibrator(s) is legitimate \citep{perrin03}, provided that the ``calibrator-science-calibrator'' observation sequence, repeated several times, rapidly samples the evaluation of the IRF \citep*{berger11}. 

To perform the calibration of the squared visibility and triple product data, \textit{SPIDAST} applies the following procedure, working at spectral channel level:
\begin{enumerate}
	\item Compute the measured IRF for each exposure, given by the ratio of the calibrator measured observable to the calibrator synthetic observable (at the same spectral resolution, for the same baseline or baseline triplet). Values and uncertainties are derived from the 4th-order approximation formulae of \citet{winzer00}. For triple product data, the approximation formulae apply separately to the real and imaginary parts of the complex ratio. Since the calibrator angular-diameter uncertainties are smaller than 10\% (see Table~\ref{TabDiamCal}), the uncertainty of a model observable is simply obtained from  its first partial derivative w.r.t. the angular diameter, multiplied by the angular diameter uncertainty.
	\item Average the IRF measurements over each OB. As weight associated with each exposure, we use the ratio of the associated weight (defined in Sect.~\ref{SubSecAver}) to the variance of the calibrator ``raw'' observable. 
	\item Determine the IRF at the mean time of each exposure obtained with the science target, from linear interpolation, or polynomial fit, on the averaged IRF measurements. Figure~\ref{FigResp} shows the temporal variation of the measured and interpolated response in squared visibility and triple product, obtained during one observing night, with three successive calibrators.
	\item Divide the observable measured on the science target for each exposure of a given OB, by each interpolated value of the IRF obtained for the same OB (using Winzer's formulae). Applying a method similar to that used with the \textit{getCal} tool of the \citet{getcal08}, a weight is associated with each calibrated ratio, which combines information on the observations of the calibrator and of the science target (science--calibrator angular separation and observation-time delay, coherence time, number of non-aberrant data). 
	\item Compute the weighted average over each OB of the ratios used in the calibration, which gives the final calibrated SPI observable (see Subsect.~\ref{SubSecFinal}). The argument of the final calibrated triple product gives the final calibrated closure phase.
\end{enumerate}	
Although the calibration procedure used by \textit{SPIDAST} is based on the standard calibration method, great care has been taken to provide reliable uncertainty estimates, thanks to the use of weights tracing the data quality at different levels of the processing. Thus, \textit{SPIDAST} produces large uncertainties on final calibrated observables obtained with input data of poor quality. Estimating reliable error bars is of crucial importance to get calibrated data, usable for the fitting process of parametric models, even on data of uneven quality (see Sect.~\ref{SecFit}).

%
   \begin{figure*}
   \centering
   \includegraphics[scale=1.0, clip=true, trim=0 0 0 0]{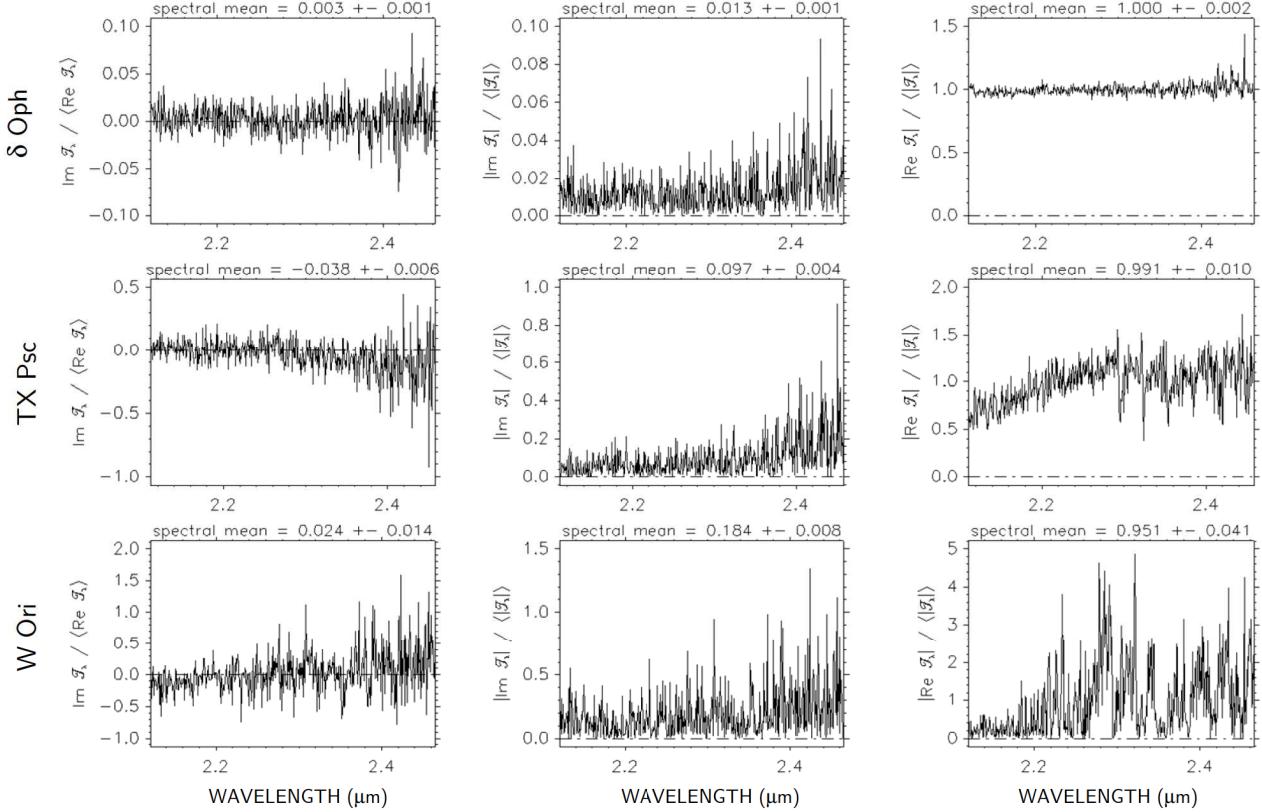}
    \caption{Left panels:~spectral variation of the imaginary part of the triple product, divided by the spectral average of the real part, for 3 science targets (the ``spectral mean'' quoted above each panel corresponds to the spectral average of the displayed quantity; for the left panels, it thus equals the tangent of the global closure phase). Central panels:~absolute value of the imaginary part of the triple product, divided by the spectral average of the modulus (for the central panel, the ``spectral mean'' equals the sinus of the CSP). Right panels:~absolute value of the real part of the triple product, divided by the spectral average of the modulus.}
             \label{CspLambda}%
    \end{figure*}    

\subsubsection{Comparing the final calibrated observables with \textit{amdlib}} \label{SubSecFinal}
 
Although the calibration routine provided with the \textit{amdlib} package is not yet officially validated \citep{malbet11}, we compute the differences in squared visibility and closure phase between \textit{SPIDAST} and \textit{amdlib}, and compare them with the associated uncertainties. 

Figure~\ref{FigDelOphCalib} shows the results obtained on the science target $\delta$~Oph, with the baseline triplet A0--D0--H0, under good and poor seeing conditions. 
The histograms of the uncertainties show that the uncertainties in squared visibility estimated with \textit{amdlib} are smaller than with \textit{SPIDAST} (whatever the seeing), while it is the contrary for the  uncertainties in closure phase. 

To explain the disagreement between the uncertainties in squared visibility, we remind that \textit{SPIDAST} takes into account instrumental and environmental conditions in the calculation of the uncertainties, while \textit{amdlib} uses only unweighted variance. Thus, we consider as underestimated the uncertainties in squared visibility produced by \textit{amdlib}.

Regarding the closure phase, it is more difficult to explain why the uncertainties from \textit{amdlib} are larger than the ones from \textit{SPIDAST}. Indeed, no details are found in \citet{malbet11}, expected to describe the \textit{amdlib} calculation. \textit{SPIDAST} computes the uncertainties on the real and imaginary parts of the calibrated triple product, and derives the uncertainties on the final closure phase from them. A deep analysis based on the source code of the calibration routine provided by \textit{amdlib} reveals that  \textit{amdlib} computes the uncertainties in the final closure phase, adding the uncertainties in closure phase of the science target and of the estimated transfer function. Since the closure phase is less stable than the triple product, the final uncertainties in closure phase handled with \textit{amdlib} are larger than with \textit{SPIDAST}. 

To evaluate the agreement between the two procedures, we compare the differences between the final calibrated SPI observables they produce, to the associated uncertainties. Since the median differences are smaller than the median uncertainties (Fig.~\ref{FigDelOphCalib}), we conclude that the agreement between the two softwares is satisfactory, \textit{SPIDAST} giving more reliable uncertainties than \textit{amdlib}.
   
\section{Measuring the deviation from circular symmetry} \label{SecCentro}

The closure phase $\psi_{123}$ couples the phases $\varphi$ of the Fourier transform of the source brightness distribution at the three spatial frequencies $f_1$, $f_2$, and $f_3=f_1+f_2$, probed by the three baselines in the following way: $\psi_{123}=\varphi(f_1)+\varphi(f_2)-\varphi(f_3)$.
In the case of a centro-symmetrical
brightness distribution  on-axis,  the phase of the Fourier transform is uniformly zero and $\psi_{123}$ is naturally equal to zero. If this source is off-axis, the phase  is linear w.r.t. the spatial frequency, so that $\varphi(f_3)=\varphi(f_1)+\varphi(f_2)$, and, subsequently,  also here $\psi_{123}$ is equal to zero.
Thus, a non-zero phase closure is a clear indication, at least qualitatively,  for a deviation from circular symmetry.

In Sect.~\ref{SubSecCsp}, we introduce a new parameter (called the \textit{centro-symmetry parameter}), based on the triple product, more sensitive to deviation from centro-symmetry than the closure phase. 

\subsection{Global closure phase} \label{SubSecClosfact}

The integration of the real and imaginary parts of the triple product $\mathscr{T}_\mathrm{true}$, over the observation spectral band $\left[\lambda_\mathrm{min}; \lambda_\mathrm{max}\right]$, leads to the global closure phase, defined by \citet{ragland06} and \citet{tatebe06} as
\begin{equation}
   \psi_\mathrm{band} =  \mathrm{atan} \ \frac{ \int^{\lambda_\mathrm{max}}_{\lambda_\mathrm{min}} \Im \mathscr{T}_\mathrm{true}\left(\lambda\right)  \: d\lambda }{ \int^{\lambda_\mathrm{max}}_{\lambda_\mathrm{min}} \Re \mathscr{T}_\mathrm{true}\left(\lambda\right)  \: d\lambda }. 
	\label{PsiFact} 
\end{equation}
A value of the global closure phase close to zero (less than \unit[1]{\degr}, for our sample) suggests a high degree of centro-symmetry of the brightness distribution, related to the observation spectral band. We compute the uncertainty on the global closure phase thanks to the direct-bootstrap method, using random sampling with replacement of the spectral dataset of the calibrated triple product. The values of the global closure phases in $K$, and their associated uncertainties, are given in the third column of Table~\ref{TabCSPclos}, for each science target.

\begin{table}
\centering
  \caption{CSP values and global closure phases, for the science targets. The last column gives the ratio of the associated SNRs.}
  \label{TabCSPclos}
\begin{tabular}{lccc}
  \hline
 Name & CSP~(\degr) & $\psi_\mathrm{band}$~(\degr) & SNR$_\mathrm{CSP}$/SNR$_\mathrm{\psi}$ \\ 
   \hline
$\delta$~Oph & 0.76(3) & 0.16(4) & 7.0 \\
$\alpha$~Car & 0.84(3) & -0.22(5) & 5.7 \\
L$_{2}$~Pup & 0.90(4) & 0.39(5) & 3.2 \\
$\beta$~Cet & 0.98(4) & 0.32(6) & 4.6 \\
$\zeta$~Ara  & 1.12(4) & -0.22(6) & 8.1 \\
$\alpha$~TrA & 1.21(5) & 0.20(7) &  8.3 \\
$\alpha$~Hya & 1.21(9) & -0.16(11) & 9.4 \\
TW~Oph & 2.94(5) & -2.91(5) & 1.1 \\
CE~Tau & 3.01(11) & -0.85(17) & 5.5 \\
$\gamma$~Hyi & 3.25(11) & 2.83(13) & 1.4 \\
$o_{1}$~CMa & 4.41(197) & 2.88(155) & 1.2 \\
$\sigma$~Lib & 5.12(20) & -1.90(33) & 4.4 \\
$\gamma$~Ret & 5.16(37) & -4.44(115) & 3.6 \\
TX~Psc & 5.34(25) & -2.17(32) & 3.4 \\
$o_{1}$~Ori & 8.10(82) & 7.48(84) & 1.1 \\
W~Ori & 10.58(46) & 1.35(79) & 13.3 \\
R~Scl & 11.63(59) & -10.63(67) & 1.2 \\
T~Cet & 27.88(51) & -28.79(41) &  0.8 \\
\hline
\end{tabular}
\end{table}

\subsection{Centro-Symmetry Parameter (CSP)} \label{SubSecCsp}

When the positive and negative values of $\Im \mathscr{T}_\mathrm{true}\left(\lambda\right)$ along the spectral domain almost mutually compensate (as seen on Fig.~\ref{CspLambda} for TX~Psc and W~Ori), the spectral integration produces a nearly null global closure phase.
As mentioned above, this can be taken as a hint for centro-symmetry. However, $\Im \mathscr{T}_\mathrm{true}\left(\lambda\right)$ is clearly non-null in some parts of the spectrum, and this is a hint for a deviation from centro-symmetry. 
To rule out the contradiction, and so as to consider this latter possibility, we introduce a new estimator, that we call the \textit{Centro-Symmetry Parameter} (CSP), similar to the global closure phase, but using instead the absolute value of $\Im \mathscr{T}_\mathrm{true}\left(\lambda\right)$ in the numerator, and the modulus of  $\mathscr{T}_\mathrm{true}\left(\lambda\right)$ in the denominator, 
     \begin{equation}
   \mathrm{CSP}  = \mathrm{asin} \ \frac{ \int^{\lambda_\mathrm{max}}_{\lambda_\mathrm{min}} \left|\Im \mathscr{T}_\mathrm{true}\left(\lambda\right)\right| \: d\lambda }{ \int^{\lambda_\mathrm{max}}_{\lambda_\mathrm{min}} \left| \mathscr{T}_\mathrm{true}\left(\lambda\right) \right| \: d\lambda }.
   \label{CspTrip} 
   \end{equation}
As for the global closure phase, a small CSP value (less than \unit[2]{\degr}, for our sample) suggests a high degree of centro-symmetry. Significantly high CSP values, as judged from their uncertainties (bootstrap method again), require to use asymmetric brightness distribution models, for the fitting on the SPI data. The values of the global CSP in the $K$-band, and their uncertainties, are  given in the second column of Table~\ref{TabCSPclos}.

%
\begin{figure}
 \includegraphics[scale=0.45, clip=true]{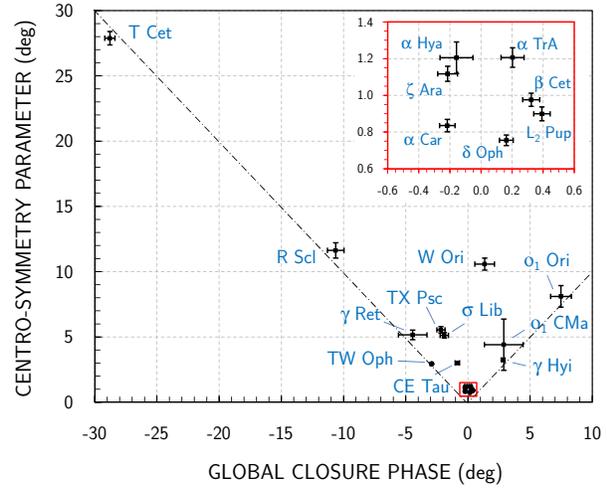}
\caption{The CSP versus $\psi_{K}$, the global closure phase in $K$, for the science targets. The red inset in the upper right corner enlarges the group of points located at the bottom, around the null value of the global closure phase, pertaining to the targets $\delta$~Oph, $\alpha$~Car, L$_{2}$~Pup, $\beta$~Cet, $\zeta$~Ara, $\alpha$~TrA, and $\alpha$~Hya. The dash-dot-dash diagonals trace the relation CSP~=~$\left|\psi_{K}\right|$.}
\label{Fig:psiCSP}  
\end{figure} 
%

To illustrate the difference between the global closure phase and the CSP, we plot in Fig.~\ref{CspLambda} three quantities, for 3 scientific targets showing a global closure phase close to zero,
\begin{enumerate}
	\item in the panels on the left is shown the imaginary part of the triple product $\Im \mathscr{T}_\mathrm{true}$, divided by the spectral mean\footnote{defined as the integral w.r.t. the wavelength, divided by the spectral bandwidth $\Delta\lambda =\lambda_\mathrm{max}-\lambda_\mathrm{min}$} of the real part $\left\langle\Re \mathscr{T}_\mathrm{true}\right\rangle$. Note that the spectral mean of this ratio is equal to the tangent of the global closure phase, as defined in Eq.~(\ref{PsiFact});
	\item in the panels forming the central column is shown the absolute value of the imaginary part of the triple product $\left|\Im \mathscr{T}_\mathrm{true}\right|$, divided by the spectral mean of the modulus $\left\langle\left|\mathscr{T}_\mathrm{true}\right|\right\rangle$. Note that the spectral mean of this ratio is equal to the sine of the global CSP, as defined in Eq.~(\ref{CspTrip});
	\item in the panels on the right is shown the absolute value of the real part of the triple product $\left|\Re \mathscr{T}_\mathrm{true}\right|$, divided by the spectral mean of the modulus $\left\langle\left|\mathscr{T}_\mathrm{true}\right|\right\rangle$. 
\end{enumerate}
The three panels on the left show three typical behaviours for $\Im \mathscr{T}_\mathrm{true}$~:
\begin{enumerate}
	\item uniformly close to zero ($\delta$~Oph);
	\item decreasing symmetrically around zero (TX~Psc);
	\item increasing symmetrically around zero (W~Ori).
\end{enumerate}
For each of these situations, the integration over the spectrum produces a nearly null result (hint for centro-symmetry). However the CSP allows to detect deviations from centro-symmetry (see Fig.~\ref{Fig:psiCSP}). 

   \begin{figure*}
	\centering
   \includegraphics[scale=0.85, clip=true, trim=0 0 0 -15]{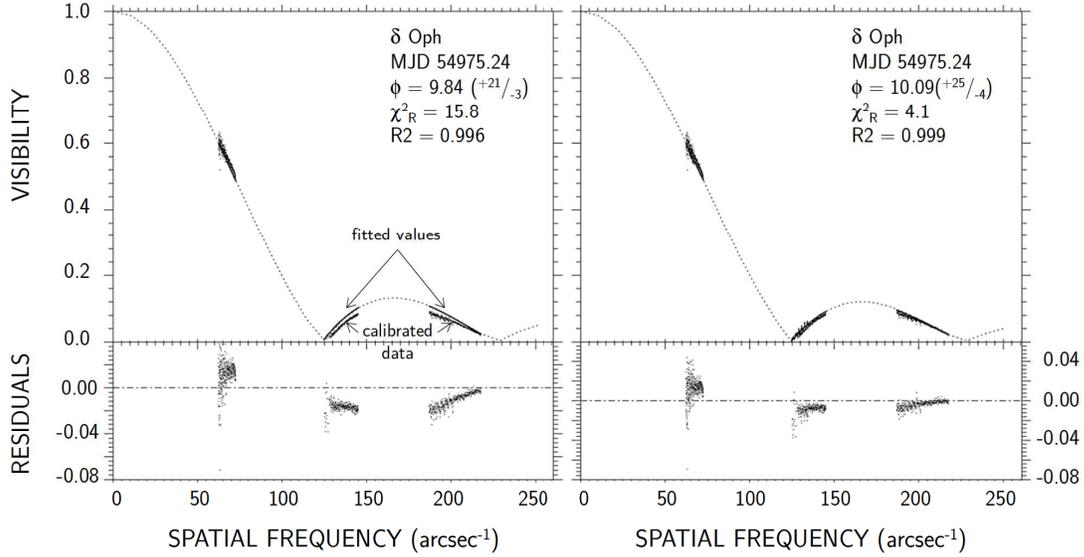}
  \caption{Fits on the visibility data of $\delta$~Oph obtained with the baselines A0--D0, D0--H0, and A0--H0, for one single OB. Left panel:~fit of the UD model; right panel:~fit of the LD-MARCS model (models in dashed lines). Calibrated data are shown without error bars, for clarity purpose. $\phi$ is the best-fit angular outer diameter (in mas), $\chi^{2}_\mathrm{R}$ the reduced chi-squared, and R2 the ajusted coefficient of determination. Bottom panels:~residuals (calibrated data - fitted values). }
              \label{FigDelOphFit}%
    \end{figure*}
%

If the real and imaginary parts of the triple product are wavelength-independent, for the observation spectral range, one can show that $\sin \mathrm{CSP} = \left|\psi_\mathrm{band}\right|$. If not, no analytical relation linking the two quantities can be derived from Eqs.~(\ref{PsiFact}) and (\ref{CspTrip}), because of the integrals in the numerators and the denominators. Figure~\ref{Fig:psiCSP} shows the CSP in $K$ displayed versus the global closure phase in the same spectral band, for each science target, choosing the OB which gives the smallest uncertainty on the CSP. 
Almost all the stars fall along the diagonals of the ($\psi_\mathrm{band}$, CSP) diagram, which trace the relation CSP~=~$\left|\psi_\mathrm{band}\right|$. One (W~Ori), however, is flagged as asymmetrical with the CSP indicator, but not with $\psi_\mathrm{band}$. 
This is precisely why we favour the CSP over $\psi_\mathrm{band}$, as discussed above.
T~Cet has the largest values for CSP and $\psi_\mathrm{band}$ as well, indicating large deviation from centro-symmetry (may be due to strong asymmetries at the wavelengths of the CO~bands or even suggestive of the presence of a binary companion). The second largest CSP value pertains to R~Scl, recently revealed as a wide binary by \citet{maercker12}, using the ALMA array at millimetric wavelengths.
In addition, the last column of Table~\ref{TabCSPclos} shows that, except for T~Cet, the SNR for the CSP is comparable or larger than the SNR for the global closure phase. 
  
\section{Fitting parametric chromatic models} \label{SecFit}

In order to interpret the final calibrated SPI observables, \textit{SPIDAST} provides a fitting routine, based on the modified gradient-expansion algorithm, very similar to the algorithm invented by \citet{levenberg44}, and improved by \citet{marquardt63}. The fit applies on any calibrated SPI measurements (to be chosen between visibility, flux, coherent flux, bispectrum, triple product, and closure phase), using a library of single-component or composite parametric chromatic models, characterised by the Fourier spectrum of their intensity distribution, and the associated first-order partial derivatives w.r.t. the model parameters. Figure~\ref{FigDelOphFit} shows examples of visibility-fitting results obtained with \textit{SPIDAST}, using two models (UD and LD-MARCS). 
This fitting routine, applied on spectro-photometric data, is used to determine the calibrator angular diameters from the MARCS + Turbospectrum synthetic spectra (see Subsect.~\ref{SubSecCaldiam}).

Note that the \textit{LITpro} software\footnote{\url{www.jmmc.fr/litpro_page.htm}}, developed by the JMMC working group ``Model fitting'', uses a set of elementary geometrical and center-to-limb darkening functions, as well \citep{bosc08}. However, it does not offer, contrary to \textit{SPIDAST}, the possibility to fit stellar-disk models with synthetic tabulated radiance data (or exitance, for fits on spectro-photometric measurements), such those produced by the MARCS + Turbospectrum code.

Since the uncertainties of the final calibrated data are not normally distributed, the covariance matrix, that comes out of the chi-squared fit, cannot be used to infer the parameter uncertainties \citep{press07}. To compute the uncertainties, \textit{SPIDAST} uses the residual-bootstrap method, described in detail in \citet*{cruzalebes10}:
\begin{enumerate}
	\item ``synthetic'' datasets are produced from random sampling with replacement of the Pearson residuals (difference between calibrated and fitted values divided by the uncertainty of the observed value), added to the initial fitted values; 
	\item fitting the model on these new datasets produces a set of chi-squared minima, following a probability distribution, from which we extract the boundaries of the confidence interval, with a given confidence level;
	\item the parameter values associated with these chi-squared boundaries give the upper and lower limits of the parameter estimates, leading to asymmetric uncertainties.
\end{enumerate}
The whole procedure allows the computing of reliable estimates of the parameter uncertainties.

\section{Conclusion} \label{SecConcl}

In the present paper, we introduce the new \textit{SPIDAST} software, developed since 2006 with the aim to reduce, calibrate, and interpret the visibility and triple product measurements obtained with the VLTI/AMBER facility. \textit{SPIDAST} contains a whole set of modules, which can be launched separately or in an automatic batch file, and summarised herafter:

\begin{enumerate}
	\item The raw data reduction used by \textit{SPIDAST} computes the weighted average of non-aberrant data, at spectral-channel level, providing estimates of the SPI observables using an automated procedure, while the method presently used with \textit{amdlib} selects the ``best'' frames according to a quality threshold determined \textit{a posteriori}, after several trials. 
  \item The wavelength calibration procedure performed by \textit{SPIDAST} provides spectral shifts following a polynomial law, tracing them at the channel level. This is done by computing the cross-correlation of the measured spectra of the calibrators with their synthetic spectra produced by the MARCS model, while \textit{amdlib} only provides a constant spectral shift over the spectrum, from the correlation with the telluric lines. 
	\item For selected calibrators not included in the calibrator catalogues, \textit{SPIDAST} provides several routines for estimation of the angular diameter with indirect methods, the most accurate being the fit of stellar-atmosphere model spectra given by MARCS on spectro-photometric data. The calibrator synthetic observables are derived using CLV functions produced by MARCS. 
	\item To obtain an accurate interferometric calibration (via an automated procedure), \textit{SPIDAST}:~(1)~divides the calibrator raw data with the associated synthetic observables in each spectral channel, which gives the instrumental response function in squared visibility and triple product;~(2)~interpolates or fits the response at the time of each exposure on the science target;~(3)~divides the science raw data with the interpolated/fitted response, which gives the science calibrated observables for each exposure;~(4)~computes their weighted average over each OB. At each processing step, the uncertainties are computed thanks to the bootstrap method applied on the weighted means. 
	\item Using the real and imaginary parts of the calibrated triple product, \textit{SPIDAST} measures the deviation from centro-symmetry of the brightness distribution of each science target in the observation spectral band, thanks to a new parameter, more sensitive to asymmetries than the global closure phase.
	\item Finally, \textit{SPIDAST} proposes a complete fitting tool, using a set of parametric and chromatic models, and accepting input tables of flux/intensity synthetic data.
\end{enumerate}
 
Such a careful calibration process of SPI data is a crucial step for their trustworthy astrophysical exploitation, which is the topic of two associated papers \citep{cruzalebes13b,cruzalebes13c}. Parameter extraction using non-linear fits of source model, as well as aperture-synthesis image reconstruction, need reliable estimates of the calibrated observables, with robust uncertainties. Our reduction, calibration, and fitting routines also apply to any other spectral datasets, including spectroscopic data. We made the \textit{SPIDAST} software public\footnote{\url{https://forge.oca.eu/trac/spidast}}: the source code of any program of our software suite can be obtained by sending an e-mail to the first author of the present paper.

\section*{Acknowledgments}

   The authors thank the ESO-Paranal VLTI team for supporting their AMBER observations, especially the night astronomers A. M\'erand, G. Montagnier, F. Patru, J.-B. Le~Bouquin, S. Rengaswamy, and W.J. de~Wit, the VLTI group coordinator S. Brillant, and the telescope and instrument operators A. and J. Cortes, A. Pino, C. Herrera, D. Castex, S. Cerda, and C. Cid. They also thank the 
Programme National de Physique Stellaire (PNPS) for supporting part of this collaborative research. 
   A.J. is grateful to B. Plez, K. Eriksson, and T. Masseron for their ongoing support on the use of 
   the MARCS code. S.S. was (partly) supported by the Austrian Science Fund through FWF project P19503-N16; A.C. is post-doctoral fellow from F.R.S.-FNRS (Belgium; grant 2.4513.11); E.P. is supported by PRODEX. 
 The data were partially reduced using the publicly available data reduction software package \textit{amdlib},
kindly provided by the Jean-Marie Mariotti Center. This study used the SIMBAD and VIZIER databases
at the CDS, Strasbourg (France), and NASAs ADS bibliographic services. 


\label{lastpage}

\end{document}